\documentclass[aip,amsmath,amssymb,reprint]{revtex4-1}

\usepackage{graphicx}
\usepackage{dcolumn}
\usepackage{bm}
\usepackage[mathlines]{lineno}
\usepackage[utf8]{inputenc}
\usepackage[T1]{fontenc}
\usepackage{mathptmx}
\usepackage{etoolbox}
\usepackage{cleveref}

\makeatletter
\def\@email#1#2{%
 \endgroup
 \patchcmd{\titleblock@produce}
  {\frontmatter@RRAPformat}
  {\frontmatter@RRAPformat{\produce@RRAP{*#1\href{mailto:#2}{#2}}}\frontmatter@RRAPformat}
  {}{}
}%
\makeatother
\begin{document}

\preprint{AIP/123-QED}

\title[]{Rethinking Pipe Flow Stability: Insights from a Meshless Global Analysis}

\author{Akash Unnikrishnan}
\author{Vinod Narayanan}%
\email{vinod@iitgn.ac.in}
\affiliation{ 
Mechanical Engineering, Indian Institute of Technology Gandhinagar, 382055, Gujarat, India}%

\homepage{https://vinod.people.iitgn.ac.in/wp/}

\date{\today}

\begin{abstract}
Despite extensive experimental evidence of turbulence in Hagen–Poiseuille flow, linear stability analysis has yet to confirm its instability. One challenge is the singularity introduced by the $1/r$ term at the pipe center ($r=0$), which complicates traditional stability approaches. In this study, we explore a global stability analysis using a meshless framework. While this approach did not recover the expected unstable modes, it revealed a new set of modes with distinct characteristics from those observed in local stability analysis. We analyze these modes and their impact on transient energy growth, demonstrating the effectiveness of the global approach in capturing localized instabilities without requiring multiple simulations. 
\end{abstract}

\maketitle

Hydrodynamic stability is fundamental to fluid dynamics and offers insight into the transition from laminar to turbulent flow. Beyond its theoretical significance, it has practical applications in engineering, meteorology, and industrial fluid systems. This transition is governed by the amplification of small perturbations, a phenomenon elegantly captured by Linear Stability Theory (LST). The pioneering works of Helmholtz \citep{helmholtz1868xliii}, Kelvin \citep{kelvin1871stratified}, Rayleigh \citep{rayleigh1882investigation}, and Reynolds \citep{reynolds1883iii} laid the groundwork for understanding hydrodynamic instability. Reynolds' seminal pipe flow experiments established the critical Reynolds number beyond which turbulence emerges, a finding that remains central to modern stability studies. The development of the Orr-Sommerfeld equations \citep{drazin1961discontinuous} further refined theoretical and experimental investigations, with significant contributions from Tollmien \citep{tollmien1936general}, Schlichting \citep{schlichting1949boundary}, and Squire \citep{squire1933stability}.  

Hagen-Poiseuille flow, first described by Hagen \citep{hagen1839ueber} and Poiseuille \citep{poiseuille1840recherches}, represents the steady, axisymmetric flow of an incompressible, Newtonian fluid through a cylindrical pipe. Despite experimental evidence of turbulence, linear stability analysis has consistently failed to identify any unstable eigenmodes. Reynolds' experiments demonstrated that, under carefully controlled perturbations, stable laminar flow can be maintained up to a Reynolds number, \(Re = 120,000\), while stronger perturbations trigger transition at much lower Reynolds numbers, typically near \(Re \approx 2000\).  

Theoretical studies have confirmed that Hagen-Poiseuille flow remains linearly stable for all Reynolds numbers \citep{salwen1980linear, meseguer2003linearized}. The transition to turbulence is thus attributed to nonlinear mechanisms and transient growth of perturbations. One challenge in conducting global stability analysis for pipe flow is the singularity at the pipe center (\(r = 0\)) due to the \(1/r\) term in the governing equations. This singularity complicates numerical approaches and is often cited as a reason for the absence of unstable eigenmodes in linear stability studies \citep{meseguer2003linearized,Schmid_Henningson_2001}. 
In this work, we employ a meshless framework \cite{shahane2021high,unnikrishnan2022shear,unnikrishnan2024taylor} developed in the Cartesian coordinates for global stability analysis \cite{unnikrishnan2024high} to circumvent this singularity and explore the stability characteristics of Hagen-Poiseuille flow.


\begin{figure}
    \centering
    \includegraphics[width=0.5\linewidth]{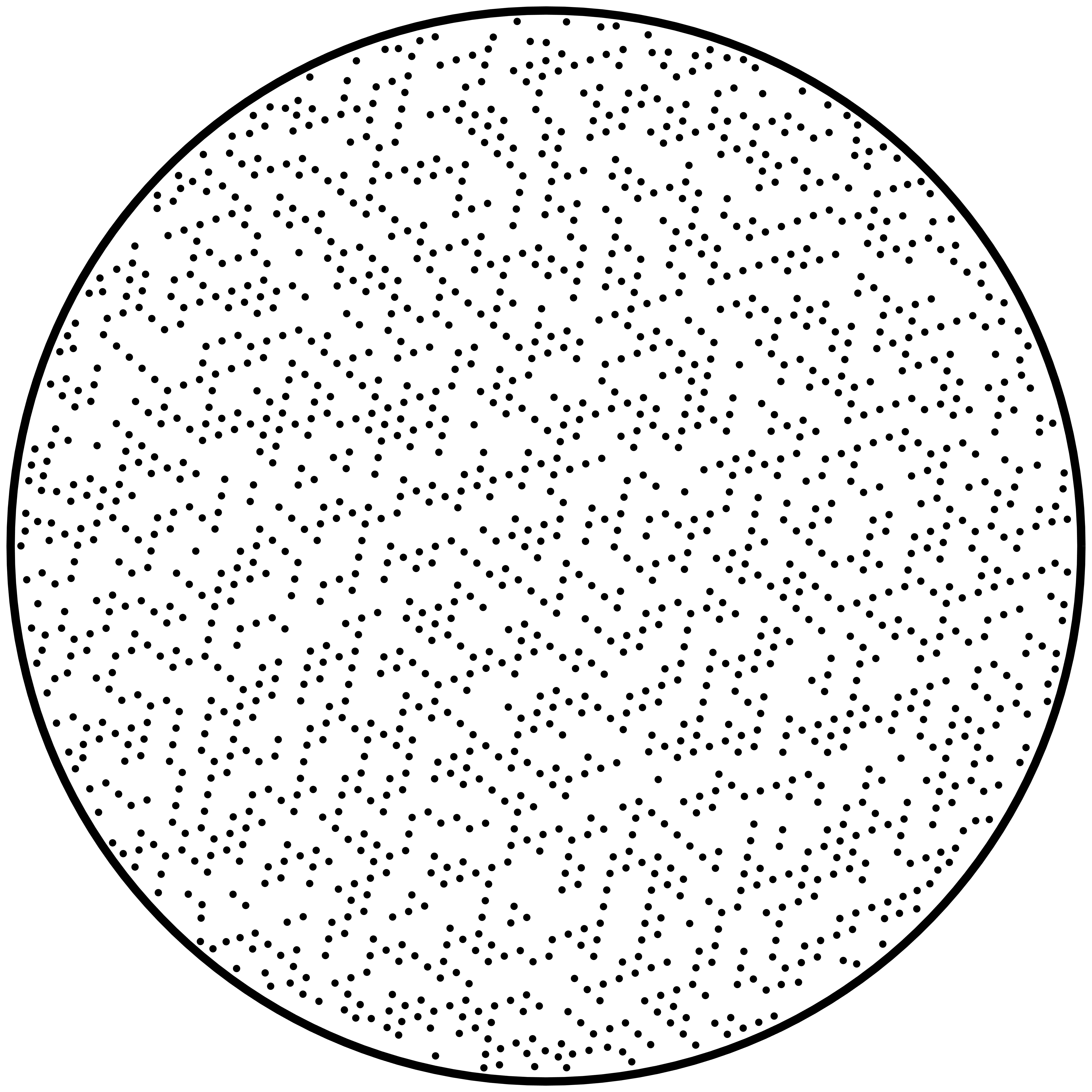}
    \caption{Sample point distribution generated using GMSH\cite{gmsh}.}
    \label{fig:point_distribution}
\end{figure}

\begin{figure*}
    \centering
    \includegraphics[width=0.8\linewidth]{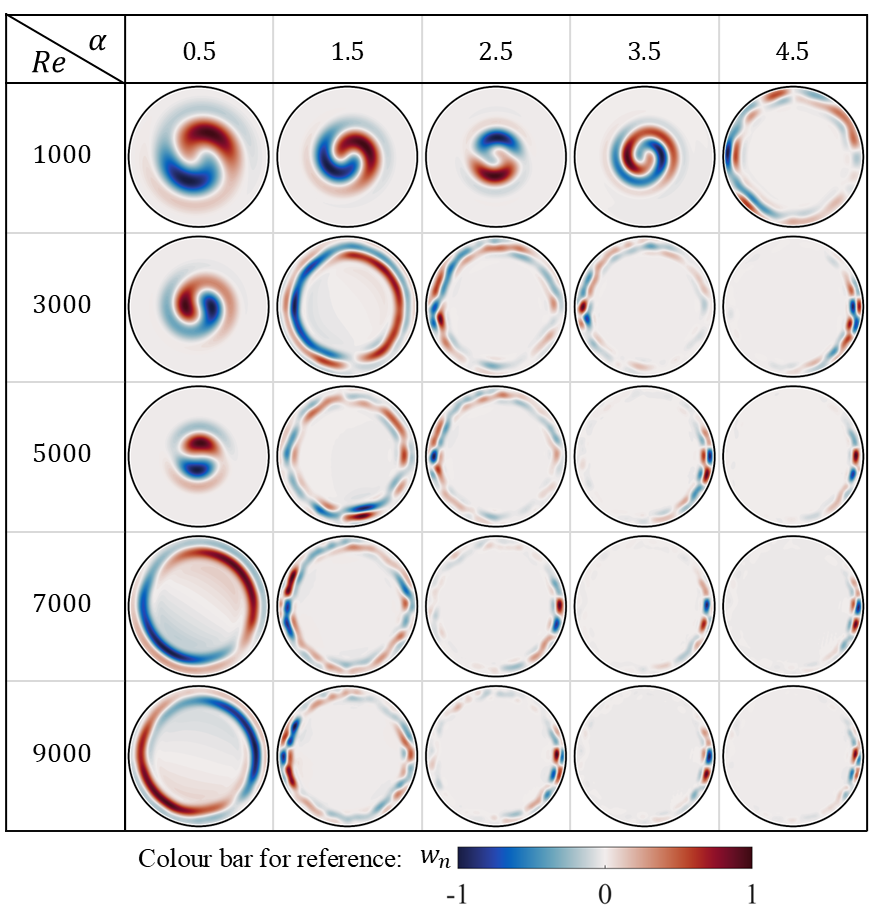}
    \caption{Most unstable mode (with largest growth rate) from global stability analysis for different Reynolds number ($Re$) and streamwise wavenumber ($\alpha$).}
    \label{fig:modes_5x5}
\end{figure*}

\begin{figure*}
    \centering
    \includegraphics[width=0.8\linewidth]{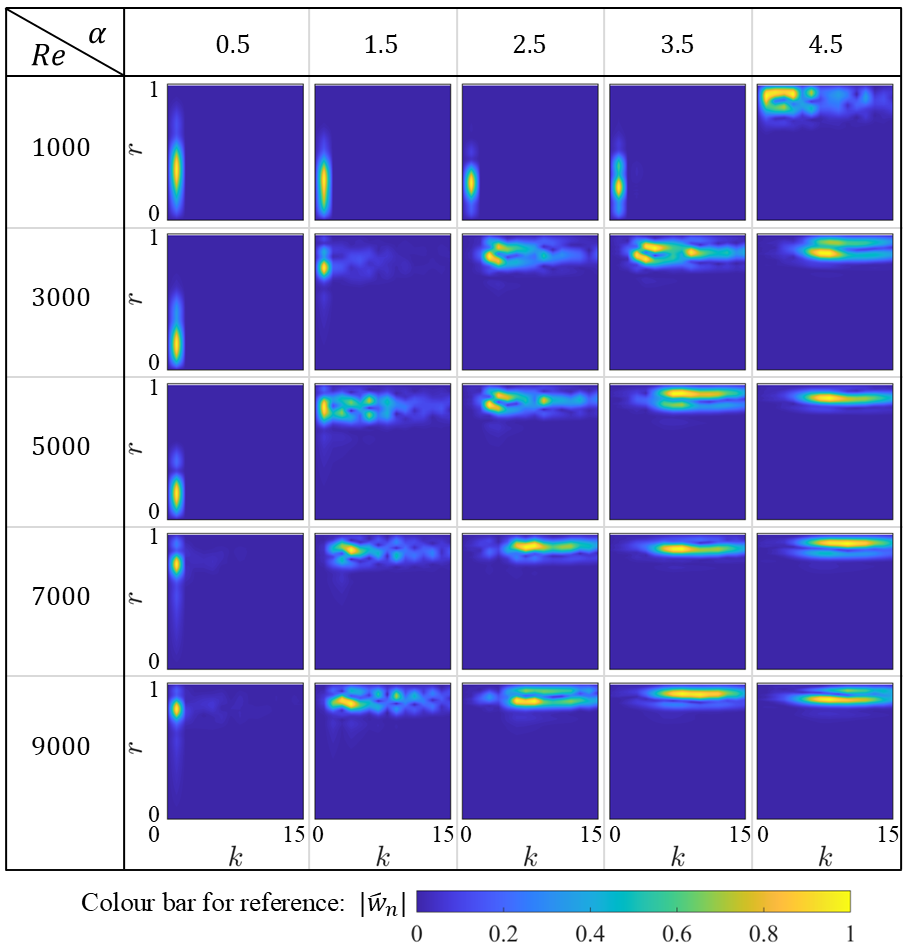}
    \caption{A 2 dimensional Fast Fourier Transform of the most unstable mode corresponding to the modes presented in \cref{fig:modes_5x5} from global stability analysis for different Reynolds number ($Re$) and streamwise wavenumber ($\alpha$). Here x-axis scales from 0 to 15 wavenumbers and y-axis scales from 0 to 1 radial distance from center of pipe in each figure.}
    \label{fig:spectrogram_5x5}
\end{figure*}

The governing equations for global linear stability analysis and the corresponding eigenvalue problem have been established in prior works \citep{alizard2007,theofilis2003advances,unnikrishnan2024high}. We briefly summarize them here and refer readers to the aforementioned references for detailed derivations. Perturbations are assumed to be of the normal mode form 
\[
q(x,y,z,t) = \hat{q}(x,y)\exp{\iota(\alpha z - \omega t)},
\] 
where \(\alpha\) is the streamwise wavenumber and \(\omega = \omega_r + i \omega_i\) is the complex eigenvalue, with its imaginary part indicating the growth rate. The resulting linearized equations are given by:
\begin{subequations}
\begin{align}
0 &= D_x \hat{u} + D_y \hat{v} + i\alpha \hat{w}, \\
i\omega \hat{u} &= D_x \hat{p} + \mathcal{L} \hat{u}, \\
i\omega \hat{v} &= D_y \hat{p} + \mathcal{L} \hat{v}, \\
i\omega \hat{w} &= i\alpha \hat{p} + D_x W \hat{u} + D_y W \hat{v} + \mathcal{L} \hat{w},
\end{align}
\end{subequations}
where the operator \(\mathcal{L} \equiv iW\alpha - \frac{1}{Re}(D_x^2 + D_y^2 - \alpha^2)\), and \(D_x = \partial/\partial x\), \(D_y = \partial/\partial y\). The base flow velocity profile in the streamwise direction is denoted by \(W(x,y)\), and \((\hat{u}, \hat{v}, \hat{w})\) and \(\hat{p}\) are the amplitudes of the perturbation velocity and pressure fields. The Reynolds number is defined as \(Re = U_{\text{max}} R/\nu\), based on the maximum velocity of the base flow \(U_{\text{max}}\), characteristic length \(R\), and kinematic viscosity \(\nu\).

We discretize the above system using a meshless collocation method based on radial basis functions (RBF) with polynomial augmentation of degree 5\citep{shahane2021high}, which provides spectral-like accuracy while eliminating the requirement for structured grids. This approach enables the computation of accurate high-order spatial derivatives, which are essential for both stability analysis and transient growth studies. This method has previously been validated for stability computations in complex geometries \citep{unnikrishnan2024high}. The collocation points are distributed over a representative cross-section of the domain, with no-slip boundary conditions imposed at solid walls and periodicity assumed in the axial direction. The resulting generalized eigenvalue problem is solved using an Arnoldi-based iterative solver, which efficiently extracts the dominant eigenmodes governing instability.

To further investigate transient amplification, we examine the linearized Navier–Stokes system in the time domain. Energy growth in a flow is driven purely by linear mechanisms, as nonlinear terms in the Navier-Stokes equations conserve energy\cite{hristova2002transient}. For a perturbation state vector \(\bm{q}' = [u,v,w,p]^T\), the evolution equation is written as
\[
\frac{\partial \bm{q}'}{\partial t} = \mathcal{A} \bm{q}',
\]
where \(\mathcal{A}\) is the discrete linear operator derived from the governing equations. The solution can be formally expressed as
\begin{subequations}
\begin{align}
\bm{q}(t) &= \bm{q}_0 e^{\mathcal{A} t}, \\
\bm{q}(t) &= \bm{q}_0 \mathbf{T}^{-1} e^{\boldsymbol{\Lambda} t} \mathbf{T},
\end{align}
\end{subequations}
where \(\boldsymbol{\Lambda}\) and \(\mathbf{T}\) denote the eigenvalues and eigenvectors of \(\mathcal{A}\), respectively.

To quantify perturbation growth, the energy norm\cite{schmid1994optimal} is defined as 
\[
E(\hat{\bm{q}}) = \int_{x} \int_{y} \int_{z} (|u|^2 + |v|^2 + |w|^2)\,dz\,dy\,dx.
\]
Assuming periodicity in the streamwise direction and integrating over one wavelength, this simplifies to
\[
E(\hat{\bm{q}}) = C_1 \int_{x} \int_{y} (|u|^2 + |v|^2 + |w|^2)\,dx\,dy,
\]
where \(C_1\) is a constant. In the meshless framework, this integral is approximated using quadrature weights computed for scattered nodes using the method of \citet{reeger2018numerical}, which is well-suited for bounded smooth surfaces.

The maximum energy amplification over time is defined as
\[
G(t) = \sup_{\bm{q}(0) \neq 0} \frac{||\bm{q}(t)||_E^2}{||\bm{q}(0)||_E^2} = ||e^{\mathcal{A} t}||_E^2,
\]
where \(||\cdot||_E\) denotes the energy norm. Since the operator \(\mathcal{A}\) is typically non-normal, its eigenvectors are not orthogonal, and transient growth may result even in stable systems. In such cases, singular value decomposition (SVD) of \(e^{\mathcal{A} t}\) is employed to determine the largest possible energy amplification. The dominant singular value \(\sigma_{\max}\) corresponds to the maximum amplification \(G(t)\), and its associated singular vectors represent the optimal initial perturbation and its evolution \citep{hristova2002transient,schmid1994optimal}.


The scattered points used in this study were generated from an unstructured triangular grid created using GMSH \citep{gmsh}. A sample distribution of these points is shown in \cref{fig:point_distribution}.

A local temporal stability analysis was carried out to identify the most unstable eigenmodes over a range of Reynolds numbers from 1000 to 10000 and streamwise wavenumbers from 0 to 10. Consistent with previous studies, all identified modes were found to be stable. The most unstable modes across various axial wavenumbers and Reynolds numbers are shown in \cref{fig:modes_5x5} as contours of normalized streamwise velocity ($w_n$). At low Reynolds numbers and low axial wavenumbers, the azimuthal mode with wavenumber one was consistently found to be the most unstable. As the Reynolds number increases, this dominant azimuthal mode remains unchanged but transitions in structure—from a bulk-flow-dominated mode to one concentrated near the wall. A similar transition is observed with increasing axial wavenumber at low Reynolds numbers: the unstable mode shifts from bulk-dominated to wall-dominated behavior.

These trends are further confirmed through a two-dimensional Fast Fourier Transform (FFT) of the modes, shown in \cref{fig:spectrogram_5x5}. In the spatial FFT spectrum, vertical ridges (local peaks in the contour) indicate dominance of specific axial wavenumbers in the unstable modes, while horizontal ridges suggest a broader distribution of wavenumbers.

As noted earlier, all identified modes are linearly stable. Previous studies have proposed that transition may arise due to nonlinear and linear interactions among these non-orthogonal eigenmodes, leading to growth in the total perturbation kinetic energy.

\begin{figure}
    \centering
    \includegraphics[width=0.9\linewidth]{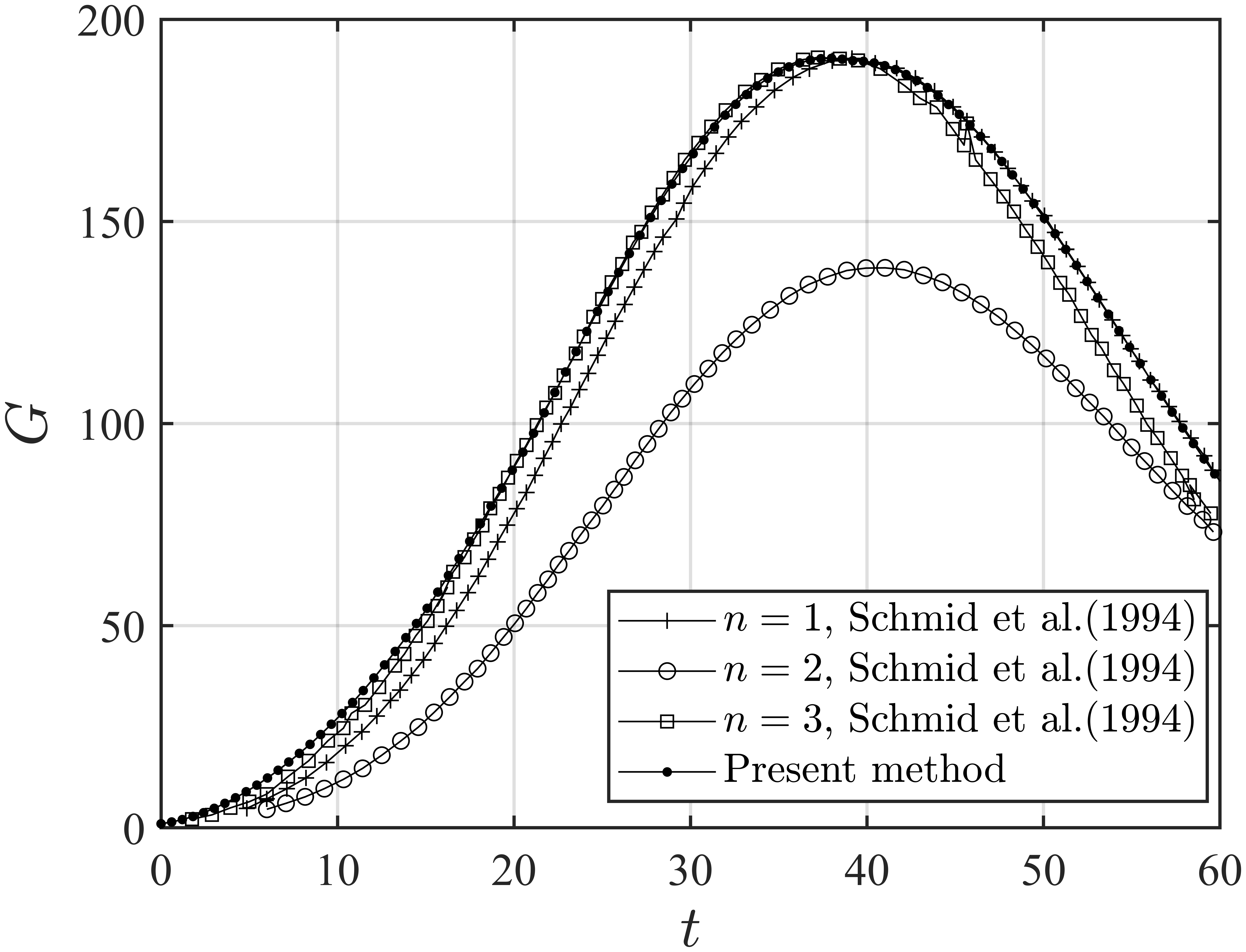}
    \caption{Transient growth curves at $Re = 3000$ and $\alpha = 1.0$ from the present meshless method compared with \citet{schmid1994optimal}}
    \label{fig:transient_growth_validation}
\end{figure}

\begin{figure}
    \centering
    \includegraphics[width=0.8\linewidth]{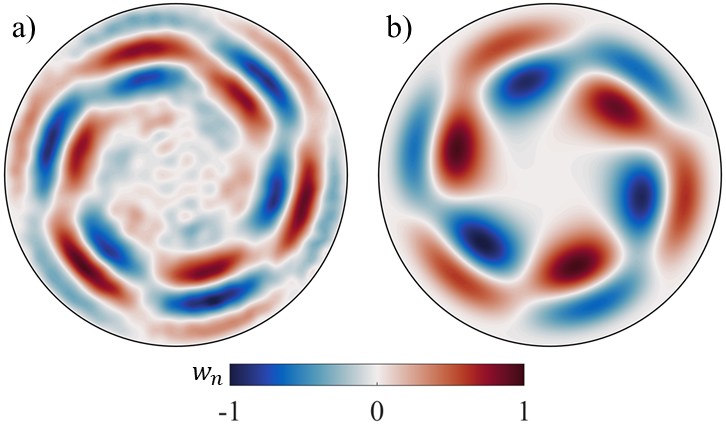}
    \caption{The optimal (a) perturbation and (b) response corresponding to the maximum growth at a $Re = 3000$ and $\alpha = 1.0$.}
    \label{fig:Re3000al1_optimum}
\end{figure}

The global method of stability analysis, although it yields a series of complex modes, results in transient growth behavior similar to that observed in the local analysis. However, the global method leads to a complete overlap of the individual growth curves for different azimuthal modes, as shown in \cref{fig:transient_growth_validation}. \citet{schmid1994optimal} previously demonstrated that each azimuthal mode exhibits a distinct growth rate. The optimal perturbation leading to such a growth is therefore confined to the corresponding azimuthal wavenumber. In contrast, the global approach allows the optimal perturbation to be a combination of multiple azimuthal wavenumbers distributed across the cross-section of the domain.

The mode corresponding to maximum transient growth for the case with Reynolds number 3000 and streamwise wavenumber 1 is shown in \cref{fig:Re3000al1_optimum}. Both the initial perturbation and the resulting response, represented as contours of streamwise velocity, correspond to an azimuthal wavenumber of 3. This observation is consistent with the findings of \citet{schmid1994optimal}, based on the perturbation kinetic energy growth curves presented in \cref{fig:transient_growth_validation}.

\begin{figure}
    \centering
    \includegraphics[width=0.9\linewidth]{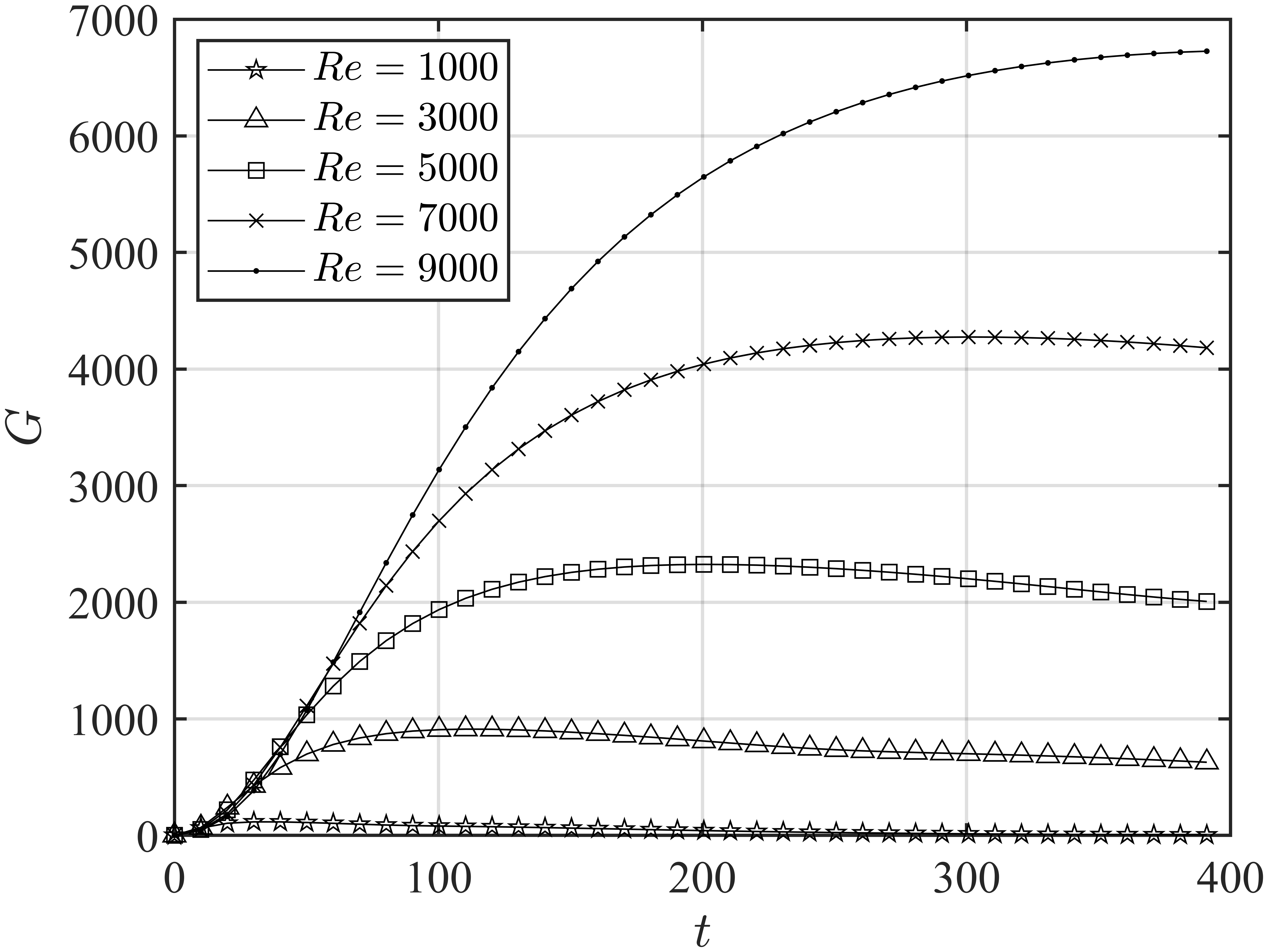}
    \caption{Transient growth curves at $Re = 1000$, $Re = 3000$, $Re = 5000$, $Re = 7000$, $Re = 9000$ and $\alpha = 0.0$ from the present meshless method compared.}
    \label{fig:transient_growth_alpha0}
\end{figure}

\begin{figure}
    \centering
    \includegraphics[width=0.8\linewidth]{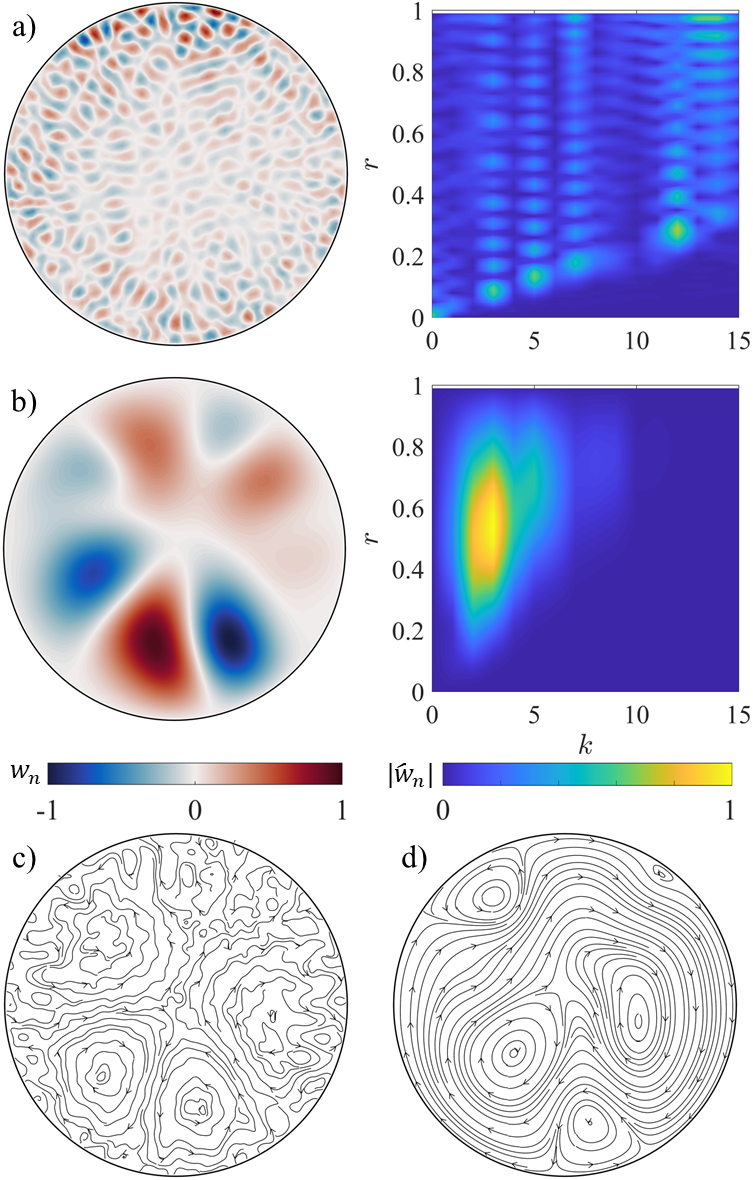}
    \caption{The optimal (a) perturbation and (b) response, represented with normalized streamwise velocity and its spatial spectrum, corresponding to the maximum growth at $\alpha = 0.0$ for all the Reynolds number tested. The optimal (c) perturbation and (d) response represented by streamlines in cross-sectional plane. }
    \label{fig:Re3000al0_optimum}
\end{figure}

The maximum energy growth was found to occur at an axial wavenumber of zero, corresponding to streak formation in the base flow. \Cref{fig:transient_growth_alpha0} shows the growth curves for different Reynolds numbers. Interestingly, the optimal perturbation and the resulting response remain qualitatively similar across the entire range of Reynolds numbers studied. \Cref{fig:Re3000al0_optimum} displays the contours of the optimal perturbation and response, with normalized streamwise velocity used as the field variable. The spatial spectra of these modes are shown alongside the contours, revealing that the amplitude of the normalized wavenumbers is distributed across a range of azimuthal wavenumbers ($k$).

\begin{figure}
    \centering
    \includegraphics[width=0.8\linewidth]{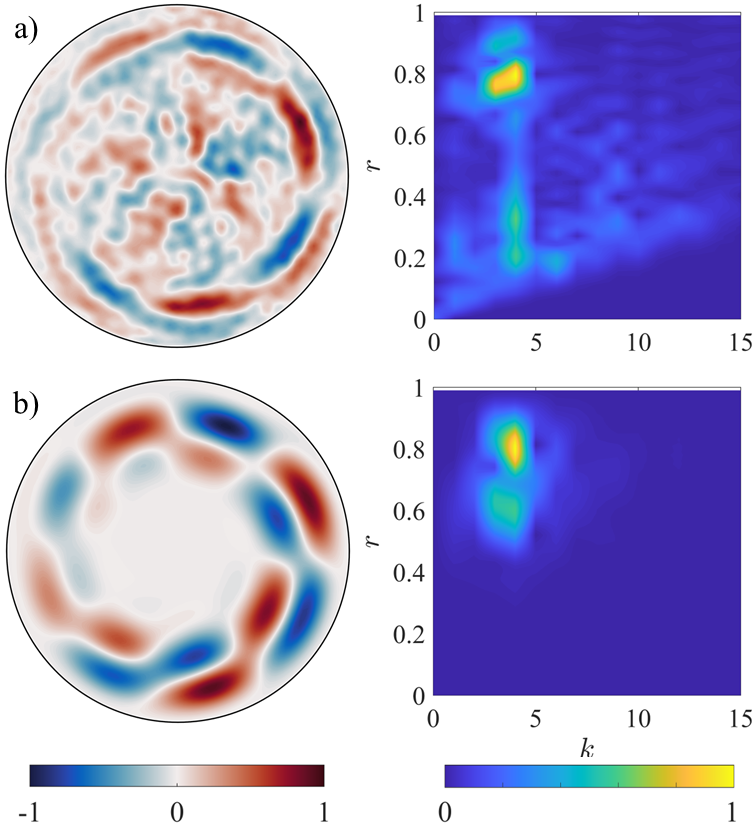}
    \caption{The optimal (a) perturbation and (b) response, represented with normalised streamwise velocity and its spatial spectrum, corresponding to the maximum growth at a $Re = 3000$ and $\alpha = 1.5$. }
    \label{fig:Re3000al3_optimum}
\end{figure}

\begin{figure}
    \centering
    \includegraphics[width=0.8\linewidth]{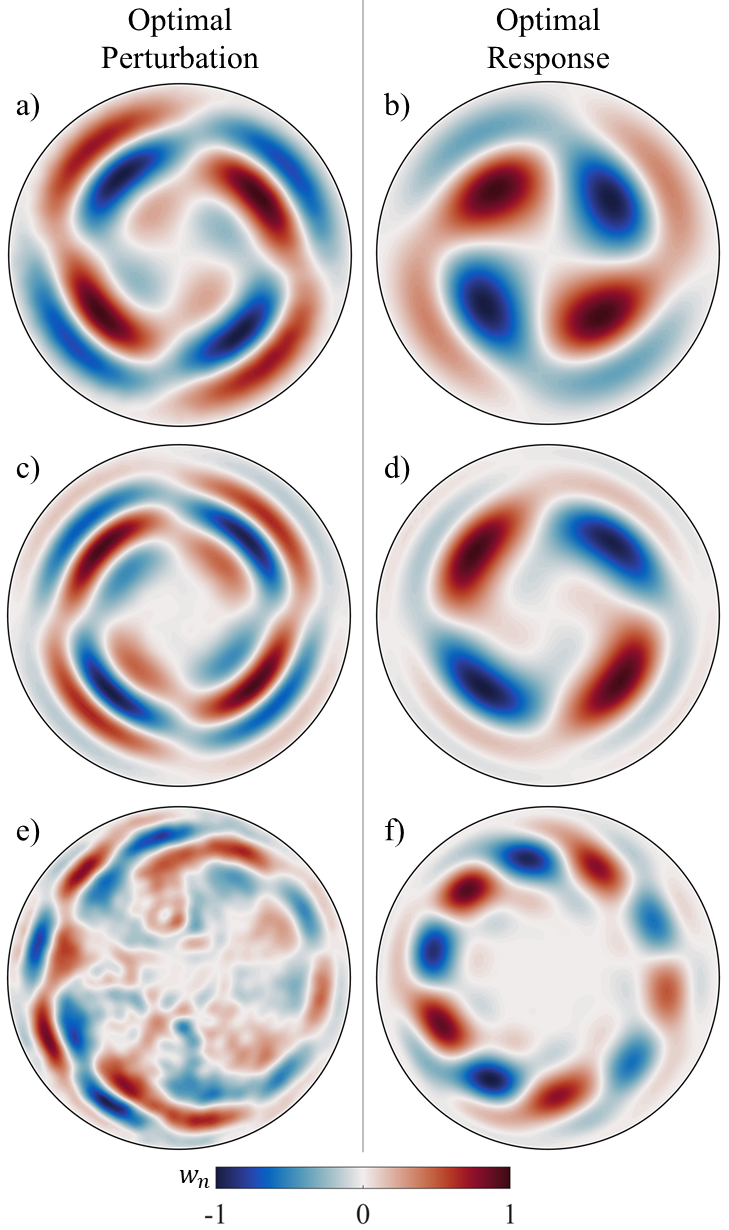}
    \caption{The optimal perturbation and response corresponding to the maximum growth at a $Re = 1000$ and (a, b) $\alpha = 1.0$; (c, d) $\alpha = 3.0$; (e, f) $\alpha = 5.0$.}
    \label{fig:Re1000al135_optimum}
\end{figure}

\begin{figure}
    \centering
    \includegraphics[width=0.8\linewidth]{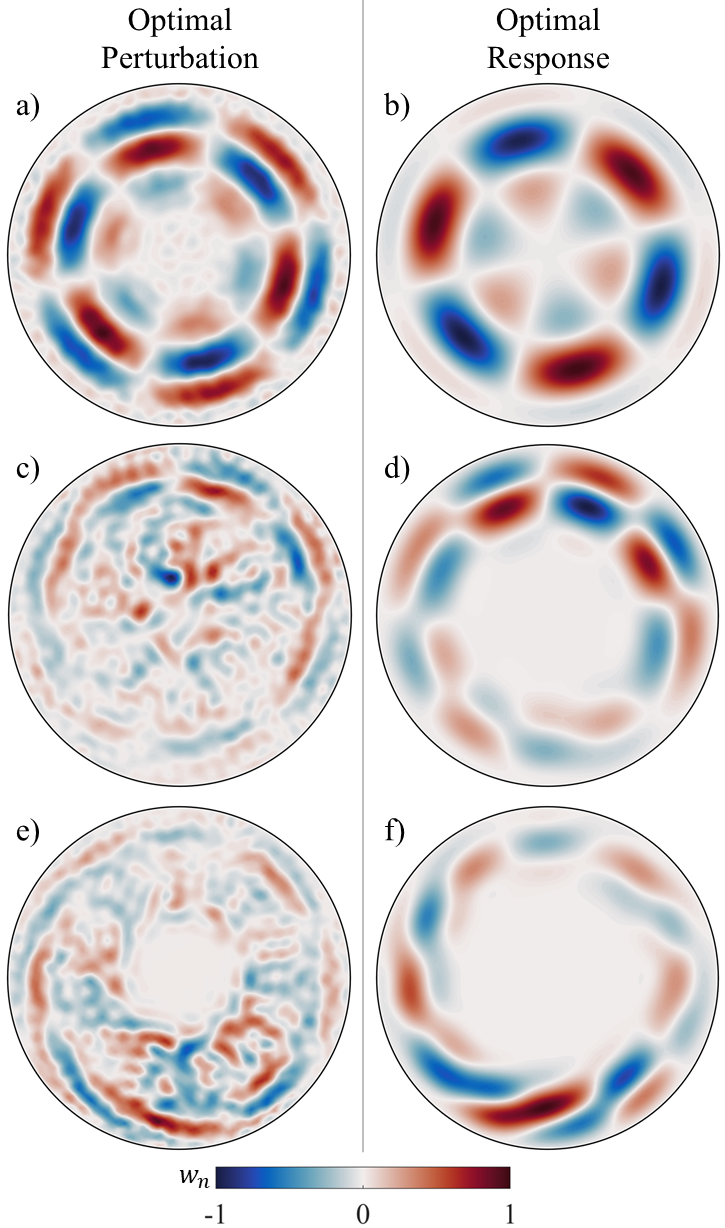}
    \caption{The optimal perturbation and response corresponding to the maximum growth at a $Re = 5000$ and (a, b) $\alpha = 0.5$; (c, d) $\alpha = 1.0$; (e, f) $\alpha = 1.5$.}
    \label{fig:Re5000al123_optimum}
\end{figure}

The optimal perturbation and response at a streamwise wavenumber of 1.5 and Reynolds number 3000 are shown in \cref{fig:Re3000al3_optimum}, along with the corresponding spatial spectrum. From the visual representation of the mode, it can be observed that the azimuthal wavenumber is not an integer but is distributed across multiple wavenumbers. The spatial spectrum displayed alongside the perturbation and response supports this observation. Although the response is centered around an azimuthal wavenumber of approximately 4, it is not strictly confined to this value but is instead spread across neighboring wavenumbers.

To further explore the difference between the local and global approaches, this method was applied at a lower Reynolds number of $Re = 1000$, for streamwise wavenumbers $\alpha = 1.0$, $3.0$, and $5.0$. The corresponding optimal perturbations and responses are shown in \cref{fig:Re1000al135_optimum}. For the lower streamwise wavenumbers ($\alpha = 1$ and $3$), the optimal modes are rotationally symmetric and primarily correspond to an azimuthal wavenumber of 2. However, for the higher streamwise wavenumber $\alpha = 5$, the optimal perturbation and response do not align with a single integer azimuthal wavenumber; instead, they involve a combination of several discrete azimuthal wavenumbers.

A similar trend is observed at a higher Reynolds number of $Re = 5000$, for $\alpha = 0.5$, $1.0$, and $1.5$. As with the $Re = 1000$ case, the optimal modes at $Re = 5000$ are not confined to a constant integer azimuthal wavenumber but are instead spread over a range of wavenumbers in spectral space.

This study began with the hypothesis that a global stability approach to pipe flow, accounting for the central singularity, might reveal temporally unstable modes. Although no unstable modes were detected across the range of Reynolds numbers (1000–10000) and streamwise wavenumbers (0–10), the most unstable modes exhibited increasingly complex structures with increasing parameters. Transient energy growth analysis showed that the global approach captures and envelops the local growth behavior, aligning with the observations of \citet{schmid1994optimal}. The optimal modes corresponding to maximum growth were found to consist of combinations of multiple azimuthal wavenumbers, emphasizing the inherently multi-modal nature of the global response. For $\alpha = 0$, the optimal perturbation and the resulting response remained qualitatively similar across the entire Reynolds number range, indicating a robust streak formation mechanism. Overall, the results demonstrate that high-order meshless methods offer a promising alternative to traditional Chebyshev–Fourier spectral approaches, enabling stability analysis in more complex geometries and revealing transient growth mechanisms not captured by local analyses.

\bibliography{aipsamp}
\end{document}